\documentclass[preprint,aps,nofootinbib]{revtex4}
\usepackage{graphicx}
\usepackage{amsmath,amssymb}
\usepackage{textcomp}
\def\ev{\mbox{eV}}

\def\AP{{\it Ann. Phys.} }

\def\CQG{{\it Class. Quantum Gravity} }

\def\JMP{{\it J. Math. Physics}}

\def\JP{{\it J. Phys.} }

\def\MPL{{\it Mod. Phys. Lett.} }

\def\NAT{{\it Nature} }

\def\PL{{\it Phys. Lett.} }
\def\PR{{\it Phys. Rev.} }
\def\PRL{{\it Phys. Rev. Lett.} }

\def\ps{\psi}

\def\bra#1{\langle{#1}|}
\def\ket#1{|{#1}\rangle}
\def\bracket#1#2{\langle{#1}|{#2}\rangle}

\def\frac#1#2{{\textstyle{{#1}\over {#2}}}}

\def\lsim{\mathrel{\rlap{\lower4pt\hbox{\hskip1pt$\sim$}}
    \raise1pt\hbox{$<$}}}
\def\gsim{\mathrel{\rlap{\lower4pt\hbox{\hskip1pt$\sim$}}
    \raise1pt\hbox{$>$}}}
\def\sqr#1#2{{\vcenter{\vbox{\hrule height.#2pt
         \hbox{\vrule width.#2pt height#1pt \kern#1pt
         \vrule width.#2pt}
         \hrule height.#2pt}}}}

\def\beq{\begin{equation}}
\def\eeq{\end{equation}}
\def\beqa{\begin{eqnarray}}
\def\eeqa{\end{eqnarray}}

\def\laq{\raise 0.4 ex \hbox{$<$}\kern -0.8 em\lower 0.62 ex\hbox{$\sim$}}
\def\gaq{\raise 0.4 ex \hbox{$>$}\kern -0.7 em\lower 0.62 ex\hbox{$\sim$}}

\begin{document}

\title{Berry Phase in the Gravitational Quantum Well and the Seiberg-Witten map.}

\author{C. Bastos}
\altaffiliation{Also at Instituto de Plasmas e Fus\~ao Nuclear, IST. Email address: cbastos@fisica.ist.utl.pt}

\author{O. Bertolami}
\altaffiliation{Also at Instituto de Plasmas e Fus\~ao Nuclear, IST. Email address: orfeu@cosmos.ist.utl.pt}

\affiliation{ Departamento de F\'\i sica, Instituto Superior T\'ecnico \\
Avenida Rovisco Pais 1, 1049-001 Lisboa, Portugal}

\vskip 0.5cm

\date{\today}

\begin{abstract}

We explicitly compute the geometrical Berry phase for the noncommutative gravitational quantum well for different SW maps. We find that they lead to different partial contributions to the Berry phase. For the most general map we obtain that $\Delta\gamma(S)\sim{\eta}^3$, in a segment S of the path in the configuration space where $\sqrt{\eta}$ is the fundamental momentum scale for the noncommutative gravitational quantum well. For the full closed path, we find, through an explicit computation, that $\gamma(C)=0$. This result is consistent with the fact that physical properties are independent of the SW map and shows that these maps do not introduce degeneracies or level crossing in the noncommutative extensions of the gravitational quantum well.

\end{abstract}

\pacs{32.80.Rm, 03.65.Ta, 11.10.Ef \hspace{2cm}Preprint DF/IST-2.2006}

\maketitle


\section{Introduction}

Noncommutative extensions of quantum mechanics (NCQM) have recently received great attention \cite{Nair,Gamboa,Duval,Ho,Horvathy,Zhang1,Acatrinei,Bertolami1,Bertolami2,Bastos}. Most of these models are based on canonical extensions of the Heisenberg algebra, actually in $d$ dimensional space extensions of the Heisenberg algebra
\beq\label{algebraNCQM}
\left[\hat q_i, \hat q_j \right] = i\theta_{ij} \hspace{0.2 cm}, \hspace{0.2 cm} \left[\hat q_i, \hat p_j \right] = i \hbar \delta_{ij} \hspace{0.2 cm},
\hspace{0.2 cm} \left[\hat p_i, \hat p_j \right] = i \eta_{ij} \hspace{0.2 cm},  \hspace{0.2 cm} i,j= 1, ... ,d
\eeq
where $\theta_{ij}$ and $\eta_{ij}$ are antisymmetric real constant $(d\times d)$ matrices and $\delta_{ij}$ is the identity matrix. By comparison between theoretical predictions for some specific noncommutative systems and experimental data, one can have bounds to these noncommutative parameters \cite{Bertolami1,Carroll}:
\beq\label{numvalues}
\theta \leq 4 \times 10^{-40} m^2 \hspace{0.5 cm},\hspace{0.5 cm} \eta \leq 1.76 \times 10^{-61} kg^2 m^2 s^{-2}.
\eeq
The main feature of this extended Heisenberg algebra is that it can be related to the standard Heisenberg algebra:
\beq\label{standalgebra}
\left[\hat R_i, \hat R_j \right] = 0 \hspace{0.2 cm}, \hspace{0.2 cm} \left[\hat R_i, \hat \Pi_j \right]
= i \hbar \delta_{ij} \hspace{0.2 cm}, \hspace{0.2 cm} \left[\hat \Pi_i, \hat \Pi_j \right] = 0 \hspace{0.2 cm},
\hspace{0.2 cm} i,j= 1, ... ,d ~,
\eeq
by a class of linear non-canonical transformations:
\beq\label{SWmap}
\hat q_i = \hat q_i \left(\hat R_j , \hat \Pi_j \right) \hspace{0.2 cm},\hspace{0.2 cm}
\hat p_i = \hat p_i \left(\hat R_j , \hat \Pi_j \right)
\eeq
which are generally referred to as Seiberg-Witten (SW) maps \cite{Seiberg}. In general, through this relation one can convert a noncommutative system into a commutative one and vice-versa. The ``new" system exhibits explicit dependence on the noncommutative parameters as well as on the particular chosen SW map. However, one should expect that the physical properties of a system do not depend on a chosen SW map, i.e. physical predictions should be independent of the particular choice of the SW map \cite{Bastos}.

Inspired in these models of NCQM, noncommutative extensions of the gravitational quantum well have been considered \cite{Bertolami1,Bertolami2,Zhang,Bertolami3}. This interest relies on the measurements of the first two quantum states of the gravitational quantum well (GQW) for ultra-cold neutrons \cite{Nesvizhevsky}. The gravitational quantum well is a system of a particle of mass m, moving in the $x'y'$ plane, subjected to the constant Earth's gravitational field, $\mathbf{g}=-g\mathbf{e_{x'}}$, with a horizontal mirror placed at $x'=0$. It is clear that this system can be studied through different SW maps. In order to study the behaviour of the GQW under different SW maps we must consider a physical feature of interest.

We point out that besides the GQW, phase space noncommutativity plays an important role also in quantum cosmology, where in the context of a Kantowski-Sachs minisuperspace model, it naturally leads to selection of states, a novel feature \cite{Bastos2}.

It is well known that quantum states submitted to adiabatic processes can acquire a geometric phase, the Berry's phase. Indeed, if the system's Hamiltonian is real then this phase can only take values 0 or $\pi$, in the end of the closed path, that is, a non-degenerated wave function must come back to itself or to minus itself. In general, the Berry phase vanishes, however when some point of degeneracy is enclosed in the loop, it can be non-trivial. One finds that Hamiltonians depending on external parameters can be affected by this phase. If, for example, the external parameter is classic, typically an external field, the Berry's phase is observed following a non-trivial loop in the space of parameters and carrying through some type of interference between the previous state and the one after completing the closed path. This might be relevant given that the next generation of experiments involving the GQW aim precisely to detect the transition and interference of states \cite{GranitII}. Given that some SW maps for phase space noncommutativity involve some momentum shift which is analogous to an external magnetic field, one may wonder whether a non-trivial Berry phase may appear in the NCGQW. If from one hand, one should not expect any non-trivial phase for systems with a real Hamiltonian that shows no crossing of levels or degeneracy, it is not clear on the other, whether this is what happens for the NCGQW. We shall see that this is actually the case of the NCGQW, irrespective of the chosen SW map.

In what follows we show, through an explicit computation, that through two distinct SW maps we obtain different phases $\Delta(S)$ for segments of a path in the configuration space, however when the full path is considered, the Berry phase is shown to vanish, irrespective of the SW map. This means that the SW map does not introduce degeneracies and level crossing in the NCGQW.


\section{The Gravitational Quantum Well}

The Hamiltonian that describes the GQW is given by
\beq \label{Hamiltonian_1}
H={{p'_x}^2\over{2m}}+{{p'_y}^2\over{2m}}+mgx'~,
\eeq
where $m$ is the neutron mass and $x'$ its height from the mirror, $p'_{x}$, $p'_{y}$ the momenta in the $x'$ and $y'$ direction, respectively \cite{Nesvizhevsky}. The associated wave function, solution of Schr\"odinger's problem, can be separated in two parts, corresponding to coordinates $x'$ or $y'$ (see e.g. \cite{Bertolami6} and references therein). In $y'$ direction, the wave function corresponds to a group of plane waves with a continuous energy spectrum. In $x$ direction, the eigenfunctions can be expressed in terms of the Airy function, $\phi(z)$:
\beq \label{Airy_1}
\psi_n(x')=A_n\phi(z)~.
\eeq
The roots of the Airy function determine the system's eigenvalues,
\beq \label{Airy_2}
E_n=-\bigg({mg^2\hbar^2\over2}\bigg)^{1/3}\alpha_n~.
\eeq
The variable $z$ is related to the height $x'$ through the expression,
\beq \label{Airy_3}
z=\bigg({2m^2g\over\hbar^2}\bigg)^{1/3}\bigg(x'-{E_n\over mg}\bigg)~.
\eeq
The normalization factor for the n-th eigenstate is given by:
\beq \label{Airy_4}
A_n=\Bigg[\bigg({\hbar^2\over 2m^2g}\bigg)^{1\over3}\int_{\alpha_n}^{+\infty}dz\phi^2(z)\Bigg]^{-{1\over2}}~.
\eeq

Let us consider a noncommutative phase space, whose algebra is given by:
\beq\label{NCalgebra}
\lbrack x,y]=i\theta\hspace{0.2cm},\hspace{0.2cm}\lbrack p_x,p_y]=i\eta\hspace{0.2cm},\hspace{0.2cm}\lbrack x_i,p_j]=i{\hbar}_{eff}\delta_{ij}\qquad i=1,2~,
\eeq
where $\hbar_{eff}=\hbar(1+{\theta\eta}/{4{\hbar}^2})$ \cite{Bertolami1}. The Hamiltonian for the noncommutative extension of the GQW, the noncommutative gravitational quantum well (NCGQW), can be obtained using a SW map that allows to re-write Hamiltonian (\ref{Hamiltonian_1}) in terms of noncommutative variables. For the most general SW map \cite{Bertolami1}
\beqa \label{SWmap1}
x'=C\left(x+{\theta\over2\hbar}p_y\right)\hspace{0.5cm},\hspace{0.5cm}y'=C\left(y-{\theta\over2\hbar}p_x\right)~,\nonumber\\
p'_x=C\left(p_x-{\eta\over2\hbar}y\right)\hspace{0.5cm},\hspace{0.5cm}p'_y=C\left(p_y+{\eta\over2\hbar}x\right)~.
\eeqa
the noncommutative Hamiltonian is given by \cite{Bertolami1}:
\beq \label{Hamiltonian_2}
H_{NC}^{(1)}={\bar{p_x}^2\over2m}+{\bar{p_y}^2\over2m}+mgx+{C\eta\over2m\hbar}(x\bar{p_y}-y\bar{p_x})+{C^2\over8m\hbar^2}\eta^2(x^2+y^2)~,
\eeq
where
\beq \label{Hamiltonian_3}
\bar{p_x}\equiv Cp_x\hspace{0.2cm},\hspace{0.2cm}\bar{p_y}\equiv Cp_y+{{m^2g\theta}\over{2\hbar}}~,
\eeq
with $C\equiv(1-\xi)^{-1}$ and $\xi\equiv \theta\eta/4\hbar^2$.

The equivalence between this set of commutation relations and the usual one for
the configuration and momentum variables, where $\xi=0$, is discussed in Ref. \cite{Bertolami2}.

Of course other SW maps could be considered. For instance, one could use instead the following transformation
\beqa \label{SWmap2}
x'=A\left(x+{\theta\over\hbar}p_y\right)\hspace{0.5cm},\hspace{0.5cm}y'=y~,\nonumber\\
p'_y=A\left(p_y-{\eta\over\hbar}x\right)\hspace{0.5cm},\hspace{0.5cm}p'_x=p_x~,
\eeqa
where $A=(1-4\xi)^{-1}$. The noncommutative Hamiltonian is now given by:
\beq \label{Hamiltonian_4}
H_{NC}^{(2)}={{p_x}^2\over2m}+{\bar{p}_{y(2)}^2\over2m}+mgx+{A\eta\over m\hbar}\bar{p}_{y(2)}x+{A^2\over2m\hbar^2}\eta^2x^2~,
\eeq
where
\beq \label{Hamiltonian_5}
\bar{p}_{y(2)}=Ap_y+{m^2g\theta\over\hbar}~.
\eeq
Thus, different SW maps give rise to different Hamiltonians. We can clearly see that the first three terms in Hamiltonians (\ref{Hamiltonian_2}) and (\ref{Hamiltonian_4}) correspond to the commutative Hamiltonian. The remaining terms are like interaction terms that, as we shall see, will affect the computation of the Berry phase.

From the identification of the first two energy eigenstates \cite{Nesvizhevsky} it is shown that the typical momentum scale is bounded by \cite{Bertolami1}:
\beq \label{nocommutation_2}
\sqrt{\eta}\lsim0.8\ \mathrm{meV}/c~.
\eeq
Assuming that $\sqrt{\theta}\lsim1\ \mbox{fm}$, the typical neutron scale, it follows that, $\xi\simeq O(10^{-24})$ and hence that the correction to Planck's constant is irrelevant.


\section{Berry Phase Computation}

Let us consider the computation of the Berry phase for the noncommutative Hamiltonians (\ref{Hamiltonian_2}) and (\ref{Hamiltonian_4}), but before that we briefly describe the context in which this phase appears. Consider a system described by a Hamiltonian, $H$. As the system evolves, the Hamiltonian changes along the trajectory and so does the state of the system. To compute the Berry phase, the starting point is Schr\"odinger's time-dependent equation
\beq \label{equation_1}
H\ket{\ps(t)} = i\hbar{d\ket{\ps(t)}\over dt}~.
\eeq
Assuming the motion is adiabatic, then the eigenvectors of the Hamiltonian evolve slowly, as follows,
\beq \label{berryphase_1}
H({\bf r}(t))\ket{n({\bf r}(t))}=E_n({\bf r}(t))\ket{n({\bf r}(t))}~,
\eeq
where ${\bf r}(t)$ is the variable describing the motion. If initially the system has a eigenstate $\ket{n({\bf r}(t))}$, then the solution of the Schr\"odinger's equation can be written as follows
\beq \label{berryphase_2}
\ket{\ps(t)}=e^{-{i\over\hbar}\int_0^t{E_n({\bf r}(t'))dt'}}e^{i\gamma_n(t)}\ket{n({\bf r}(t))}~.
\eeq
Hence, the system remains in the same eigenstate, in spite its evolution. Actually, the system acquires a dynamical and geometrical phase. This is the so-called Berry phase and can be computed by the formula \cite{Berry}:
\beq \label{berryphase_3}
\gamma_n(C)=i\int_{C}{\bracket{n({\bf r}(t))}{\nabla n({\bf r}(t))}\cdot d{\bf r}}~,
\eeq
where $C$ is a curve generated by ${\bf r}(t)$. Using the Stokes theorem one can replace the integral over $C$ by a surface integral. Through the use of a complete set of states $\sum_{m}{\ket m\bra m}=\hat{1}$, one obtains the expression,
\beq \label{berryphase_4}
\gamma_n(\partial S)=
i\int_{S}{\sum_{m\neq n}\left[{\bra n\nabla H\ket m\times\bra m\nabla H\ket n}\over{(E_n-E_m)^2} \right]\cdot{d}^2{\bf s}}~.
\eeq
In what follows we shall use Eq. (\ref{berryphase_4}) to compute the Berry phase for the GQW for the neutron and its noncommutative extension. Actually, this type of study is fairly natural in the context of cold neutron physics given that the Berry phase was first detected through the manipulation of an ultra-cold beam by polarized neutrons in a slowly varying magnetic field \cite{Bitter,Richardson}.

Given that the energy spectrum of the GQW is non-degenerated, the Berry phase should vanish. As we shall see, an explicit computation will show that the same occurs for the noncommutative extensions of the GQW and for distinct SW maps.

We first compute the Berry's phase for the commutative Hamiltonian of a particle subjected to the gravitational field. It is easy to see that
\beq \label{comutative_1}
\nabla H=\left(mg,0,0\right)~,
\eeq
since $p'_x$ and $p'_y$ are independent of coordinates $x'$ and $y'$. So,
\beqa\label{comutative_2}
\bra n\nabla H\ket m&=&mg\bracket nm=0\cr
\bra m\nabla H\ket n&=&mg\bracket mn=0~.
\eeqa
Consequently, the Berry's phase vanishes.

Let us now consider the NCGQW. From the Hamiltonian (\ref{Hamiltonian_2}), we obtain
\beq \label{berryphase_5}
\nabla H=\bigg(mg+{{C^2\eta^2}\over{4m\hbar^2}}x+{{C\eta}\over{2m\hbar}}\overline{p}_y,{{C^2\eta^2}\over{4m\hbar^2}}y-
{{C\eta}\over{2m\hbar}}\overline{p}_x,0\bigg)~;
\eeq
thus,
\beq \label{berryphase_6}
\bra n\nabla H\ket m\times\bra m\nabla H\ket n= \bigg({{C^3\eta^3}\over{8m^2\hbar^3}}\bigg)\bigg[\bra n\overline{p}_x\ket m\bra m x\ket n-\bra n x\ket m\bra m\overline{p}_x\ket n\bigg]{\bf e}_z~.
\eeq
Coordinates $y$ and $\overline{p}_y$ do not affect the computations as in $y$ direction the motion is
free and described by plane waves. As the gravitational field acts in $x$ direction, only the
coordinates $x$ and $\overline{p}_x$ will play a role in the calculation. Hence,
\beq \label{berryphase_7}
\sum_{m\neq n}{{\bra n\nabla H\ket m\times\bra m\nabla H\ket n}\over{(E_n-E_m)^2}}=\bigg({{C^3\eta^3}\over{8m^2\hbar^3}}\bigg)
\bigg[\sum_{m\neq n}{{\bra n\overline{p}_x\ket m\bra m x\ket n}\over{(E_n-E_m)^2}}{\bf e}_z-\sum_{m\neq n}{{\bra n x\ket m\bra m\overline{p}_x\ket n}\over{(E_n-E_m)^2}}{\bf e}_z\bigg]\ .
\eeq
Using $\sum_{m}{\ket m\bra m}=\hat{1}$ and that $\left[p_x,x\right]\simeq - i\hbar$, given that $\xi<<1$, one obtains
\beq \label{berryphase_8}
\sum_{m\neq n}{{\bra n\nabla H\ket m\times\bra m\nabla H\ket n}\over{(E_n-E_m)^2}}=\bigg({{-i}\over{m^2{\hbar}^2}}\bigg)\bigg({{C\eta}\over{2}}\bigg)^3\sum_{m\neq n}{{1}\over{(E_n-E_m)^2}}{\bf e}_z~,
\eeq
Substituting $E_n$ and $E_m$ for the values given by Eq. (\ref{Airy_2}):
\beq \label{berryphase_9}
\sum_{m\neq n}{{\bra n\nabla H\ket m\times\bra m\nabla H\ket n}\over{(E_n-E_m)^2}}=-i\bigg({{2}\over{m^4g^2{\hbar}^4}}\bigg)^{2/3}\bigg({{C\eta}\over{2}}\bigg)^3\sum_{m\neq n}{{1}\over{(\alpha_m-\alpha_n)^2}}{\bf e}_z,
\eeq
where $\alpha_n$ and $\alpha_m$ are the roots of the Airy function corresponding to the m-th and n-th eigenstates.
It is clear that the contribution of any state $n$ is, for the noncommutative case, non-vanishing.
To get the Berry's phase, we have to integrate this value over a closed path as in Eq. (\ref{berryphase_4}).
Thus, the following step is to choose a surface of integration adjusted to the problem of the neutron in the quantum gravitational well. It is easy to see that the only way to execute this closed path involves a transition from a state $\ket n$ to the state $\ket{n+1}$ and then back to state $\ket n$. When the neutron goes of from state $\ket n$ to state $\ket{n+1}$
one encounters a term of the type:
\beq \label{berryphase_11}
\sum_{n}{{\bra n\nabla H\ket {n+1}\times\bra {n+1}\nabla H\ket n}\over{(E_n-E_{n+1})^2}}~.
\eeq
In the following step, from state $\ket{n+1}$ to state $\ket{n}$, the
term is of the form:
\beq \label{berryphase_12}
\sum_{n}{{\bra {n+1}\nabla H\ket n\times\bra n\nabla H\ket {n+1}}\over{(E_n-E_{n+1})^2}}=-\sum_{n}{{\bra n\nabla H\ket {n+1}\times\bra {n+1}\nabla H\ket n}\over{(E_n-E_{n+1})^2}}~,
\eeq
and hence for the integrand of Eq. (\ref{berryphase_4}) one obtains
\beq \label{berryphase_13}
i{\sum_{n}{{\bra n\nabla H\ket {n+1}\times\bra {n+1}\nabla H\ket n}\over{(E_n-E_{n+1})^2}}}+i{\sum_{n}{{\bra {n+1}\nabla H\ket n\times\bra n\nabla H\ket {n+1}}\over{(E_n-E_{n+1})^2}}}=0~.
\eeq

Therefore, one concludes that the neutron does not acquire a geometric phase once it completes a closed path.
It is easy to see that this result generalizes to any number of intermediate states. This proves that the used SW map to describe the NCGQW in terms of commutative variables does not introduce degeneracies in the spectrum. To complete the proof one must show that the same is obtained for a distinct SW map.

Indeed, we consider now the Hamiltonian given by Eq. (\ref{Hamiltonian_4}). One obtains
\beq \label{berryphase_14}
\nabla H^{(2)}=\bigg(mg+{{A^2\eta^2}\over{m\hbar^2}}x+{{A\eta}\over{m\hbar}}\bar{p}_{y(2)},0,0\bigg)~.
\eeq
Thus,
\beq \label{berryphase_15}
\bra n\nabla H^{(2)}\ket m\times\bra m\nabla H^{(2)}\ket n= \stackrel{\rightarrow}{0}~.
\eeq

Once again, we obtain a vanishing Berry phase. Hence, different SW maps do lead to the same physical phase and therefore, we conclude that SW maps do not introduce degeneracies in the spectrum of the NCGQW.


\section{Discussion and Conclusions}

In this note, we have shown that the Berry phase for the noncommutative extension of the GQW vanishes likewise its commutative counterpart. Actually, we have obtained for the most general SW map that in a segment S of the configuration space, $\Delta\gamma(S)\sim{\eta}^3$, where $\sqrt{\eta}$ is the fundamental momentum scale for the NCGQW. For the full contour, $\gamma(S)=0$. This indicates that the Hamiltonian of the NCGQW is non-degenerate (besides of being obviously real). This is shown to hold for different SW maps, even though the intermediate steps of the computation yield different results. Given that the Berry phase is a physical propriety of the system it should not be affected by any set of non-canonical transformations like the SW map \cite{Bastos} provided it does not introduce degeneracies and level crossing in the system. Our explicit computation shows that this does not occur, irrespective of the SW map.

It is interesting to point out that some two-dimensional phase space noncommutative models with $\theta=const$ and $\eta=const$ can 
be described as a $3-$dimensional noncommutative model with $\eta=0$. This is the case of the 2-dimensional oscillator provided 
one introduces a momentum-dependent noncommutative, $\theta=\theta(p)$ with a ``p-monopole-form" \cite{Berard}. The case of $\eta=0$ 
has also implications for the anomalous Hall effect as it leads to a transverse displacement similar to a Berry phase \cite{Horvathy2}. 
Notice however that this does not happen for the GQW as a magnetic field like term arises only if $\eta\neq0$ \cite{Bertolami1}, 
and our results indicate that the Berry phase of the GQW vanishes.

As a final remark, we point out that a relevant issue to consider in the study of a geometrical phase concerns the transition between different states of the GQW. These transitions can be induced by variations of the gravitational field, minor impurities which may lead to changes in the relevant nuclear potential, gradients of magnetic fields, etc. It is interesting to point out that given the uncertainty principle limitation on the energy precision that can be achieved in experiments with neutrons, namely $\Delta E\sim 10^{-18}\ \ev$, the detection of ``exotic" causes to the transitions between states can be ruled out. Indeed, nonlinearities in the Schr\"odinger equation \cite{Bertolami4} and Lorentz violating terms \cite{Bertolami5} are experimentally bound to be smaller than $10^{-19}\ \ev$ \cite{Bollinger}. Violations of the Equivalence Principle by polarized objects \cite{Kobzarev} are bound to be smaller than $2\times10^{-24}\ \ev$ \cite{Romalis}. Direct spin coupling to Earth rotation and its gravitomagnetic
field are of order $10^{-19}\ \ev$ and $10^{-29}\ \ev$, respectively \cite{Bini}. A putative spin-torsion coupling
is expected to be at least $20$ orders of magnitude smaller \cite{Alimohammadi}. Hence, it is reasonable to conclude that
transitions between the GQW states, whether observed, will most likely be due to conventional effects.

\vskip 0.5cm

\centerline{\bf {Acknowledgments}}

\vskip 0.5cm

\noindent
C. Bastos would like to thank Funda\c c\~ao para a Ci\^encia e a Tecnologia (Portuguese Agency) for the financial support under
the fellowship SFRH/BD/24058/2005.

\vskip 0.5cm





\begin{thebibliography}{99}

\bibitem{Nair} V.P. Nair and A.P. Polychronakos, \PL {\bf B 505} (2001) 267.

\bibitem{Gamboa} J. Gamboa, M. Loewe and J.C. Rojas, \PR {\bf D 64} (2001) 067901.

\bibitem{Duval} C. Duval and P.A. Horvathy, \JP {\bf A 34} (2001) 10097.

\bibitem{Ho} Pei-Ming Ho and Hsien-Chung Kao, \PRL {\bf 88} (2002) 151602.

\bibitem{Horvathy} P.A. Horv\'athy, \AP (N.Y.) {\bf 299} (2002) 128.

\bibitem{Zhang1} Jian-zu Zhang, \PRL {\bf 93} (2004) 043002; Jian-zu Zhang, \PR {\bf B 584} (2004) 204.

\bibitem{Acatrinei} C. Acatrinei, Mod. Phys. Lett. {\bf A 20} (2005) 1437.

\bibitem{Bertolami1} O. Bertolami, J. G. Rosa, C. Arag\~ao, P. Castorina and D. Zappal\`a, \PR\textbf{D 72} (2005) 025010.

\bibitem{Bertolami2} O. Bertolami, J. G. Rosa, C. Arag\~ao, P. Castorina and D. Zappal\`a, \MPL\textbf{A21} (2006) 795-802.

\bibitem{Bastos} C. Bastos, O. Bertolami, N.C. Dias and J.N. Prata, {\it Weyl-Wigner Formulation of Noncommutative Quantum Mechanics}, to appear in \JMP, hep-th/0611257.

\bibitem{Carroll} S.M. Carroll, J.A. Harvey, V.A. Kosteleck\'y, C.D. Lane, T. Okamoto, Phys. Rev. Lett. {\bf 87} (2001) 141601.

\bibitem{Seiberg} N. Seiberg and E. Witten, JHEP {\bf 9909} (1999) 032.

\bibitem{Zhang} Jian-zu Zhang, hep-th/0508164 (2006).

\bibitem{Bertolami3} O. Bertolami and J. G. Rosa, \PL\textbf{B633} (2006) 111.

\bibitem{Nesvizhevsky} V. V. Nesvizhesky \emph{et al.}, \NAT\textbf{415} (2002) 297; \PR\textbf{D 67} (2003) 102002; {\it Eur. Phys. J.} 
\textbf{C40} (2005) 479.

\bibitem{Bastos2} C. Bastos, O. Bertolami, N. Dias and J. Prata, to appear in \PR {\bf D}, 0712.4122 [gr-qc].

\bibitem{GranitII} http://lpsc.in2p3.Sr/congres/granit06/index.php

\bibitem{Bertolami6} O. Bertolami and F.M. Nunes, \CQG\textbf{20} (2003) L61.

\bibitem{Berry} M. V. Berry, {\it{Proc R. Soc. London}} \textbf{A 392} (1984) 45.

\bibitem{Bitter} T. Bitter and D. Dubbers, \PRL\textbf{59} (1987) 251.

\bibitem{Richardson} D. J. Richardson, A. I. Kilvington, K. Green,  S. K. Lamoreaux, \PRL\textbf{61} (1988) 2030.

\bibitem{Berard} A. B\'erard and H. Mohrbach, \PR\textbf{D 69} (2004) 127701.

\bibitem{Horvathy2} P. Horv\'athy, \PL\textbf{A359} (2006) 705.

\bibitem{Bertolami4} O. Bertolami, \PL\textbf{A 154} (1991) 225.

\bibitem{Bertolami5} O. Bertolami and J. G. Rosa, \PR\textbf{D 71} (2005) 097901.

\bibitem{Bollinger} J. J. Bollinger et al., \PRL\textbf{63} (1989) 1031.

\bibitem{Kobzarev} Yu. J. Kobzarev and L. B. Okun, {\it{JEPT}} \textbf{16} (1963) 1343.

\bibitem{Romalis} M. V. Romalis, W. C. Griffiths, J. P. Jacobs and E. N. Fortron, \PRL\textbf{86} (2001) 2505.

\bibitem{Bini} D. Bini, C. Cherubini and B. Mashhoom, \CQG\textbf{21} (2004) 3893.

\bibitem{Alimohammadi} M. Alimohammadi and A. Shariati, {\it Eur. Phys. J.} {\bf C 21} (2001) 193.


\end{thebibliography}
\end{document}